\documentclass[hyper]{prop2015}
\usepackage[english]{babel}

\usepackage[mathscr]{eucal}
\usepackage{amsmath,amsfonts,amssymb,amsthm,amsxtra}
\usepackage{bbm}

\renewcommand{\tilde}{\widetilde}
\renewcommand{\hat}{\widehat}

\renewcommand{\simeq}{\cong}
\newcommand{\bref}[1]{\textbf{\ref{#1}}}
\newcommand{\Hom}{\mathop{\mathrm{Hom}}}

\newcommand{\p}[1]{|#1|}
\newcommand{\gh}[1]{\mathrm{gh}(#1)}

\newcommand{\dd}{\partial}
\renewcommand{\d}{\partial}
\renewcommand{\dh}{\mathrm{d_h}}

\renewcommand{\geq}{\,{\geqslant}\,}

\newcommand{\binner}[2]{{\langle}\kern-4.15pt{\langle}#1{,}\,#2{\rangle}\kern-4.15pt{\rangle}}
\newcommand{\commut}[2]{[#1{,}\,#2]}

\newcommand{\half}{\mathchoice{\ffrac{1}{2}}{\frac{1}{2}}{\frac{1}{2}}{\frac{1}{2}}}
\newcommand{\ffrac}[2]{\raisebox{.5pt}{\footnotesize$\displaystyle\frac{#1}{#2}$}\kern1pt}

\newcommand{\dl}[1]{\mathchoice{\ffrac{\dd}{\dd #1}}{\frac{\dd}{\dd #1}}{\ffrac{\dd}{\dd #1}}{\ffrac{\dd}{\dd #1}}}
\newcommand{\dr}[1]{\ffrac{{\overset{\leftarrow}{\partial}}}{ \partial #1}}

\newcommand{\Liealg}{\mathfrak} 
\newcommand{\algg}{\Liealg{g}}

\newcommand{\fR}{\mathbbm{R}}

\newcommand{\fZ}{\mathbbm{Z}}

 \def\cE{\mathcal{E}}

\renewcommand{\proof}{{\bf Proof.}~}

\newcommand{\mto}{\mapsto}
\newcommand{\beq}{\begin{eqnarray}}
 \newcommand{\eeq}{\end{eqnarray}}
 \newcommand{\be}{\begin{array}}
 \newcommand{\ee}{\end{array}}

\newcommand{\Id}{\mathrm{Id}}
\newcommand{\uHom}{\underline{\mathrm{Hom}}}

\newcommand{\md}{\mathrm{d}}

\newcommand{\cP}{{\mathcal{P}}}
\newcommand{\cC}{{\mathcal C}}

\newcommand{\cJ}{\mathcal{J}}

\newcommand{\cF}{{\mathcal{F}}}

\newcommand{\R}{{\mathbbm R}}
\newcommand{\Z}{{\mathbbm Z}}

\def\g{{\mathfrak g}}

\def\a{\alpha}

\def\e{\epsilon}

\def\de{\delta}

\def\sst{\scriptscriptstyle}

\def\BG-Poincare{Barnich:2009jy}
\def\Fedosov-book{Fedosov:1996fu}

\theoremstyle{plain}
\newtheorem{theorem}{Theorem}[section]

\newtheorem{proposition}[theorem]{Proposition}
\newtheorem{prop}[theorem]{Proposition}
\newtheorem{cor}[theorem]{Corollary}
\newtheorem{rem}[theorem]{Remark}
\newtheorem{example}[theorem]{Example}
\theoremstyle{definition}
\newtheorem{deff}[theorem]{Definition}

\category{Proceedings}
\title{Gauge PDE and AKSZ-type Sigma Models}
\subtitle{\href{http://www.maths.dur.ac.uk/lms/109/index.html}{EPSRC/LMS Durham Symposium on Higher Structures in M-Theory}}
\author[M.~Grigoriev]{Maxim Grigoriev\inst{a,}\footnote{Corresponding author e-mail:~\href{mailto:grig@lpi.ru}{\textsf{grig@lpi.ru}}}}
\author[A.~Kotov]{Alexei Kotov\inst{b}}
\address[1]{Tamm Department of Theoretical Physics, Lebedev Physics Institute, Leninsky ave. 53, Moscow 119991, Russia and Institute for Theoretical and Mathematical Physics, Lomonosov Moscow State University, Moscow 119991, Russia}
\address[2]{Faculty of Science, University of Hradec Kralove, Rokitanskeho 62, Hradec Kralove 50003, Czech Republic}
\shortauthors{M. Grigoriev and A. Kotov}

\begin{abstract}
A gauge PDE is a natural notion which arises by abstracting what physicists call a local gauge field theory defined in terms of BV-BRST differential (not necessarily Lagrangian). We study supergeometry of gauge PDEs paying particular attention to globally well-defined definitions and equivalences of such objects. We demonstrate that a natural geometrical language to work with gauge PDEs is that of $Q$-bundles. In particular, we demonstrate that any gauge PDE can be embedded into a super-jet bundle of the $Q$-bundle. This gives a globally well-defined version of the so-called parent formulation. In the case of reparameterization-invariant systems, the parent formulation takes the form of an AKSZ-type sigma model with an infinite-dimensional target space.
\end{abstract}

\begin{acknowledgements}
We are grateful to A.\,Verbovetsky for the collaboration at the early stage of this project. M.G. is also grateful to H.\,Khudaverdian and T.\,Voronov for illuminating discussions. The work of M.G. has been supported by the Russian Science Foundation grant 18-72-10123. The research of A.K. was supported by the grant no. 18-00496S of the Czech Science Foundation.
\end{acknowledgements}

\shortabstract

\begin{document}
\maketitle

\section{Introduction}

Ideas and methods originating from gauge theories play a prominent role in both modern theoretical physics and mathematics. By all means this applies to Batalin--Vilkovisky (BV) quantization~\cite{Batalin:1981jr,Batalin:1983wj}, which allows to reformulate physical questions as cohomological problems, giving them an invariant meaning. Moreover, the structures originating in BV approach are now actively studied from a pure mathematical perspective \cite{Stasheff:1997fe,Getzler:1994yd,Khudaverdian:2000zt,beilinson2004chiral,Voronov:2009nr,Pantev:1111.3209,Felder:2012kn,Calaque:2014:926-947,paugam2014towards,costello2016factorization,DiBrino:1801.03770}.

In the original context of local gauge field theories, besides the quantization itself, BV approach offers a rigorous framework~\cite{DuboisViolette:1985jb,Barnich:1993vg,Barnich:1994db,Piguet:1995er,Barnich:2000zw} to address questions such as deformations, anomalies, global symmetries, conserved charges etc. This is achieved by defining the BV formalism\footnote{Here and below we use use the term BV formulation to refer to its natural generalization to theories defined at the level of equations of motion. In the BV approach this corresponds to forgetting the odd-symplectic structure and working in terms of the BV-BRST differential in place of the master action. The generalization is rather natural and was put forward in~\cite{Barnich:2004cr} (earlier somewhat implicit discussions can be found in~\cite{Brandt:1997iu}). Interesting generalizations to the case of non-Lagrangian systems with Lagrange anchor was put forward in~\cite{Lyakhovich:2004xd,Kazinski:2005eb,Kaparulin:2011xy}. Alternative partially Lagrangian systems were discussed in~\cite{Sharapov:2016qne,Grigoriev:2016wmk}.} in terms of suitable jet-bundles (see e.g.~\cite{Anderson:1989aa,Dickey:1991xa,Olver:1993baa,vinogradov2001cohomological,Krasilshchik:2010sst}), which approximate the infinite-dimensional geometry of the space of field histories. In this framework it becomes clear  that a local gauge field theory can be considered as a geometrical object generalizing  partial differential equations (PDE). More specifically, as seen from BV perspective, a local gauge field  theory can be considered as a PDE equipped with extra structures. What is more important, the natural equivalence~\footnote{In the present context the notion of equivalence was proposed in~\cite{Dresse:1990dj} in the context of Lagrangian BV formulation and then extended in~\cite{Barnich:2004cr} (see also~\cite{Barnich:2010sw,Grigoriev:2012xg}) to the BV at the level of equations of motion. The recent discussion from the $L_\infty$-algebra perspective can be found in~\cite{Jurco:2018sby}.} of gauge PDEs  differs from that of usual PDEs, making them an interesting objects to study even on their own.

Although in the conventional approach the BV formulation of a given system is constructed in term of its equations of motion, gauge symmetries, and (higher order) reducibility relations, it even turns out that it is useful to define gauge theory in the BV language. This point of view was explicitly put forward in~\cite{Barnich:2004cr} (see also~\cite{Barnich:2010sw,Barnich:2015tma}) and is supported by a number of examples including models of string field theory~\cite{Thorn:1986qj,Bochicchio:1986zj,Bochicchio:1986bd}, higher spin gauge theories~\cite{Vasiliev:1988sa,Vasiliev:1992av,Vasiliev:2003ev}, and topological systems~\cite{Alexandrov:1995kv} (see also~\cite{Cattaneo:1999fm,Grigoriev:1999qz,Batalin:2001fc,Park:2000au,Roytenberg:2002nu,Kazinski:2005eb}), where the theory is build from the very start in the form of its BV formulation. Further examples are gauge theories of boundary values in the context of the AdS/CFT correspondence, which immediately arise in BV form if one starts from the BV formulation in the bulk~\cite{Bekaert:2012vt,Bekaert:2013zya,Bekaert:2017bpy,Grigoriev:2018wrx}.

Defining a theory in BV terms from the outset naturally leads to a more general class of systems than one would arrive  by applying BV procedure to a given theory defined in terms of a Lagrangian or equations of motion. This motivates introducing a notion of gauge PDE as a system defined from the very beginning in terms of BV-BRST differential. One of the goals of the present work is to give a general and geometrical version of the definition of gauge PDE.

Just like usual PDEs gauge PDEs can also be defined intrinsically irrespective to the embedding into a jet-bundle. Although such an approach is known in the literature and has proved useful in applications a globally well-defined geometrical definition was missing. Filling this gap is another goal of the present work.

It was observed~\cite{Barnich:2005ru,Barnich:2010sw} that the appropriate geometrical setup for gauge PDEs is provided by so called AKSZ sigma models or more specifically their generalizations to not necessarily Lagrangian systems. AKSZ sigma models were originally proposed~\cite{Alexandrov:1995kv} as nice BV formulations for some topological theories. A somewhat similar unfolded approach~\cite{Vasiliev:1988xc,Vasiliev:2005zu} has been also independently developed in the context of higher spin gauge theories. It turns out that at least locally any reparameterization invariant gauge PDE can be brought to AKSZ form at the price of allowing infinite-dimensional target space. This is achieved by employing a so-called parent construction, proposed in~\cite{Barnich:2010sw} (see also~\cite{Barnich:2004cr} for the linear case and~\cite{Grigoriev:2010ic,Grigoriev:2012xg} for Lagrangian systems)  in local setting. In this work we propose a globally defined version of this construction and elaborate on its properties. In so doing we actively employ so-called $Q$-bundles~\cite{Kotov:2007nr}, which provide a proper geometrical setup for AKSZ models and their generalizations.

The paper is organized as follows. In Section~\bref{sec:prelim} we introduce basic notions such as $Q$-manifolds and their equivalence, define gauge PDEs through a usual jet-bundle BV formulation. In the main section, Section~\bref{sec:Q-bundle}, we give a new more flexible and invariant definition of gauge PDEs and their equivalence in terms of $Q$-bundles, define a generalized parent construction is these terms and prove that for a good gauge PDE its parent formulation is an equivalence. Finally we discuss possible applications and further perspectives.

\section{Preliminaries}\label{sec:prelim}

\subsection{Q-manifolds and their equivalences}

\begin{deff}\cite{Schwarz:1992gs}
A $Q$-manifold is a $\Z$-graded supermanifold equipped with a degree one vector field $Q$, which satisfies the nilpotency condition $[Q,Q]=2Q^2=0$.
\end{deff}

\begin{example} $T[1]X$, the shifted tangent bundle to a smooth manifold $X$, whose  (non-negatively graded) algebra of functions is isomorphic to the
algebra of differential forms $\Omega^\bullet (X)$, is a $Q$-manifold with the $Q$-field being equal to the de Rham operator. A non-negatively graded $Q$-manifold is also called an $NQ$-manifold.
\end{example}

Let us enumerate some other important examples of $Q$-manifolds. Hereafter we shall use the standard convention for the shifted parity: if $L=\bigoplus_{i\in\Z}L^i$
is a $\mathbbm{Z}$-graded vector space, then $L[p]$ is another $\mathbbm{Z}$-graded vector space, defined such that $L[p]^i = L^{i+p}$. A degree $p$ linear operator, acting between two $\mathbbm{Z}$-graded
vector spaces $L_1$ and $L_2$, can be also regarded as a degree preserving linear map $L_1\to L_2 [p]$. It is easy to see that $\left(L[p]\right)[q]$ is canonically
isomorphic to $L[p+q]$. Given two $\Z$-graded vector spaces $L_1$ and $L_2$, their direct sum $L_1\oplus L_2$ and algebraic tensor product $L_1\otimes L_2$
are naturally $\Z$-graded vector spaces.
The graded dual vector space $L^*=\bigoplus_{i\in\Z}\left(L^*\right)^i$ is defined such that $\left(L^*\right)^i=\left(L^{-i}\right)^*$,
which implies that the bi-linear pairing $L\otimes L^*\to \R$ has degree $0$.
\smallskip
\begin{enumerate}[i)]
\item Given a vector space $\g$, a $Q$-structure on the corresponding graded supermanifold $\g[1]$, the algebra of functions of which $\cF(\g[1])=\Lambda^\bullet (\g^*)$,
 is in one-to-one correspondence with a Lie algebra structure on $\g$, such that the Chevalley--Eilenberg cochain differential $\md_{\rm CE}$ is the $Q$-field.
\item In general, a $Q$-field on $A[1]$, the shifted vector bundle on $X$, is in one-to-one correspondence with a Lie algebroid structure on $A$ \cite{vaintrob:algebroids}.
This example extends the notion of a Lie algebra and the shifted tangent bundle, simultaneously.
\item Another generalization of a Lie algebra is an $L_\infty$-algebra \cite{Lada:1992wc}. Let $L$ be a $\Z$-graded vector space, $L[1]$ be a graded manifold, whose algebra of functions is $\cF(L[1])=\hat{S}(L[1]^*)$, where $\hat{S}(V^*)$, the formal completion of the algebra of graded symmetric polynomials on a graded vector space $V$, is defined as
\begin{equation}
\hat{S}(V^*)=\varprojlim S^{\le k} (V)*=\varprojlim \left( S(V^*)/ S^{\ge k+1}(V^*)\right)\,.
\end{equation}
An $L_\infty$-structure on $L$ is in one-to-one correspondence with a pointed $Q$-structure on $L[1]$, that is, a $Q$-field vanishing at the origin (cf.~\cite{Kontsevich:1997vb}).
The Taylor power expansion of $Q$ at $0$
gives us a series of $k$-linear operators $S^k(L[1])\to L[2]$ or, equivalently, using a natural isomorphism of graded vector spaces $S^k(L[1])\simeq \Lambda^k (L)[k]$,
a series of skew-symmetric (in the graded sense) $k$-linear operators $l_k\colon \Lambda^k L\to L[2-k]$, which satisfy the compatibility conditions determined by the corresponding
$Q$-field. In particular, the nilpotency condition $[Q,Q]=0$ implies that $l_1$ is a degree one differential, while $l_2$ is a degree zero skew-symmetric bilinear $L$-valued form
(a pre-Lie structure on $L$), the Jacobi identity of which is satisfied up to the homotopy term $l_1\circ l_3+l_3\circ l_1$, etc.
\end{enumerate}

\begin{deff}There is a category $\mathsf{QMan}$, whose objects are $Q$-manifolds and morphisms are degree preserving maps $\phi\colon (M_1, Q_1)\to (M_2,Q_2)$, which are
compatible with the $Q$-structures; the latter means that the pull-back map on functions $\phi^*\colon \cF(M_2)\to \cF(M_1)$
is a chain map:
\beq
Q_1\circ \phi^* = \phi^* \circ Q_2\,.
\eeq
\end{deff}
The category of $Q$-manifolds is supplied with the unit object $(pt, 0)$ and the direct product of $Q$-manifolds $(M_1, Q_1)$ and $(M_2, Q_2)$, such that
the underlying $\Z$-graded $Q$-manifold is the direct product $M_1\times M_2$ and the $Q$-structure $Q_{12}$ is uniquely determined by the property
$Q_{12}(fh)=Q_1(f)h+(-1)^{\deg f}f Q_2 (h)$ for any $f\in \cF(M_1)$ and $h\in\cF(M_2)$. The latter makes $\mathsf{QMan}$ into a symmetric monoidal category (cf.~\cite{0387984038}). In addition to the hom-set $\Hom (M_1, M_2)$, consisting of homomorphisms of $Q$-manifolds, later on being referred to as $Q$-maps, there exists the internal hom $\uHom (M_1, M_2)$, an object in a (normally) larger category of super $Q$-spaces, which is characterized by the property
\beq
\Hom (M\times M_1, M_2)\simeq \Hom \left(M, \uHom (M_1, M_2)\right)\,
\eeq
  for any $M\in \mathsf{QMan}$. In particular, when $M_1$ has a compact base (i.e. a compact zero-degree part), the internal hom is a well-defined (generally) infinite-dimensional
$Q$-manifold \cite{Bonavolonta:1304.0394}. The graded super space $\uHom (M_1, M_2)$ together with the $Q$-structure is called
the space of super maps between $(M_1, Q_1)$ and $(M_2, Q_2)$, also known in physical literature as super fields.
Its coordinate description is more transparent: let $M_1$ and $M_2$
be flat graded supermanifolds with coordinates $(x^a, \xi^\mu)$ and $u^i$, respectively, such that $\deg (x^a)=0$, $\deg (\xi^\mu)=l_\mu$, $l_\mu \neq 0$, and $\deg (u^i)=m_i$. Then
a super map $M_1\to M_2$ is expressed as follows:
\beq
u^i = \sum\limits_{\mu_1, \ldots, \mu_r} \xi^{\mu_1}\cdots \xi^{\mu_r} \phi_{\mu_1\ldots\mu_r}(x)\,,
\eeq
where $\phi_{\mu_1\cdots\mu_r}(x)$ are smooth functions of $x$, declared to be of the degree $m_i-\sum\limits_{j=1}^r l_{\mu_j}$.
Then a morphism of graded super manifolds is a super map characterized by the property that all non-vanishing coefficients
$\phi_{\mu_1\cdots\mu_r}(x)$ have degree $0$. The $Q$ field on the space of super maps is induced in the standard way
by the left- and right- infinitesimal actions of the corresponding $Q$-fields.

\begin{deff}\cite{Kotov:2007nr}
A $Q$-bundle is a fibered bundle in the category $\mathsf{QMan}$, that is, a locally trivial $\Z$-graded bundle $\pi\colon E\to M$
over a $Q$-manifold $M$, supplied with a total $Q$-structure, such that the projection map is a $Q$-morphism. A $Q$-section is
a $Q$-morphism $\sigma\colon M\to E$, such that $\pi\circ\sigma =\Id$.
\end{deff}
\begin{example}\label{ex:tangent_map}
Let $\pi_{\sst X}\colon E\to X$ be a fibered bundle over a smooth manifold, then $\pi=\md \pi_{\sst X}\colon (T[1]E, \md_E)\to (T[1]X, \md_X)$
is a $Q$-bundle.
\end{example}

\begin{rem} $\md \pi_{\sst X}$ from Example \ref{ex:tangent_map} is locally trivial as a $Q$-bundle, which means that it is locally isomorphic to the
product of the shifted tangent bundles of the base and the fiber. However, not every $Q$-bundle is locally trivial; this depends on the base
of the $Q$-manifold and, in the case of an infinite-dimensional fiber, also on the class of functions that we regard as smooth. Later on we will see examples of infinite-dimensional
$Q$-bundles over $T[1]X$ (for a  finite-dimensional $X$), which are not locally trivial in the above sense.
\end{rem}

\begin{deff}\label{def:contractible_Q-manifold}
Let $V$ be a $\Z$-graded vector space (for simplicity, we assume that the grading is bounded either from above or from below).
We shall call $\left(T[1]V, \md_V\right)$ a contractible $Q$-manifold.
\end{deff}

By the definition, a contractible $Q$-manifold possesses a homogeneous coordinate system $(w^\a, v^\a)$, such that $Qw^\a = v^\a$.

\begin{deff}\label{def:Q-reduction}
A $Q$-bundle $\lambda\colon M'\to M$ is called an equivalent reduction of $Q$-manifolds (or an equivalence $Q$-reduction) if $\lambda$ admits
a global $Q$-section $\sigma\colon M\to M'$
and a local trivialization (as a $Q$-bundle) over some open cover $\{U_i\}$ of $M$ with a contractible  $Q$-fiber $T[1]V$ for some $V$.
\end{deff}

\begin{rem} Given an equivalent reduction of $Q$-mani\-folds, we can always find a local trivialization which is compatible with the section $\sigma$ in the sense that $\sigma|_{U_i}$ coincides with the canonical inclusion $U_i\hookrightarrow U_i\times T[1]V$. Indeed, let $(u^j,w^\a, v^\a)$
be a system of adapted coordinates on $\lambda^{-1}(U)$ over a coordinate chart $(U,u^j)$, such that $Q'u^j=Qu^j=Q^j (u)$ and
$Q'w^\a=v^\a$. Assume that $\sigma |_U$ is determined by equations $\hat{w}^a=\hat{v}^\a=0$, where
$\hat{w}^a=w^\a -f^a(u)$, $\hat{v}^a=v^\a-h^a (u)$.  Taking into account that $\sigma$ is a $Q$-morphism, we immediately
obtain $Q'f^a (u) =h^a (u)$ and thus $Q'\hat{w}^\a =\hat{v}^a$. A new adapted coordinate system $(u^j, \hat{w}^\a, \hat{v}^\a)$
is obviously compatible with $\sigma$.
\end{rem}

\begin{deff}\label{def:Q-equivalence}
The minimal equivalence relation generated by the equivalence $Q$-reduction is called an equivalence of $Q$-manifolds.
\end{deff}

\begin{proposition}\label{prop:equal_cohomology} Equivalent $Q$-manifolds have the same $Q$-cohomology in all natural $Q$-complexes.
\end{proposition}

\noindent\proof
It is sufficient to verify this statement for an equivalent $Q$-reduction $\lambda\colon M'\to M$.
In order to do this, we use a trivialization of $\lambda$ over an open cover of $M$, which is compatible with the
section $\sigma$, and then the associated \u{C}zech hypercomplex. In particular, let $\cJ$ be the sheaf of functions
on $M'$, vanishing on the image of $\sigma$. Then $\cJ$ is acyclic over any $U_i$ and thus it is globally acyclic.
Given that $\cF(M')=\lambda^*\cF(M)\oplus \cJ$ as a $Q$-complex, this proves that $H_Q^k (\cF(M'))\simeq H_Q^k (\cF(M)) $
for all $k$. A similar argument can be applied to all tensor fields, viewed as a $Q$-complex with respect to the Lie derivative along $Q$.
\hfill $\square$
\smallskip

\begin{rem} The property (\ref{prop:equal_cohomology}) will remain true if we omit in Definition \ref{def:Q-equivalence}
 the condition that $M'\to M$ is a $Q$-bundle; more precisely,
we may consider an equivalence relation, generated by all pairs of $M\subset M'$ of embedded $Q$-submanifolds together with the property
that $M'$ admits an open cover $U'_i$, such that $U'_i\simeq \left(U'_i\cap M\right)\times T[1]V$ for all $i$. We use the $Q$-bundle structure in Definition \ref{def:Q-equivalence} for some technical convenience.
\end{rem}

The language of trivial and equivalent $Q$-manifolds turns out to be a useful tool in various cases.
For instance, let us illustrate how equivalent reductions often arise in applications. Given a $Q$-manifold $(M,Q)$ suppose that locally we succeeded to identify independent functions $\{w^a\}$ such that $\{w^a,Qw^a\}$ are also independent functions. It then turns out that the surface defined by $\{w^a=0,Qw^a=0\}$ is a $Q$-submanifold. Moreover, for $M$ finite-dimensional $M$ is locally a trivial $Q$-bundle over the surface. In general one is to check whether the bundle is locally trivial. 

A typical example where in this way one indeed produces an equivalent reduction is a linear  $Q$-manifold $(M,Q)$, associated to a complex $(M,\delta)$, i.e. $M$ is a graded vector space and $\delta$ is a nilpotent linear operator of degree $1$.  Let on $M$ there is an additional degree such that it is bounded from below and $\delta$ decomposes into a sum of homogeneous components $\delta=\delta_{-1}+\delta_0+\ldots$. It follows $\delta_{-1}$ is again a differential and let $\bar M$ be its cohomology and $\{w_a,v_a,u_i\}$ a basis in $M$ such that $\delta_{-1}v_a=w_a$ and $\delta_{-1} u_i=0$. If $u^i,w^a,v^a$ are coordinate functions dual to basis elements
$\{u_i,w_a,v_a\}$ then $Qw^a=v^a+\ldots$ and moreover $w^a,Qw^a,u^i$ is also a legitimate coordinate system. A subspace $w^a=0,Qw^a=0$ is a linear $Q$-submanifold which corresponds to an equivalent complex $(\bar M,\tilde \delta)$, where $\tilde \delta$ is a differential induced by $\delta$ in the cohomology of $\delta_{-1}$.
The above considerations are known~\cite{Barnich:2004cr} in the context of gauge theories and can be seen as a super-geometrical interpretation of the spectral sequence technique applied to the filtered complex.

More general example of equivalent $Q$-reduction arise in $L_\infty$ algebras.
Consider an $L$-infinity algebra $L$ together with the corresponding $Q$-structure on $L[1]$, denoted as $Q_L$. Let $\bar{L}$ be the cohomology of the complex $(L, l_1)$. 
Assume that $\imath\colon \bar{L}\hookrightarrow L$ is a `harmonic-type' embedding of the cohomology into the whole complex as a graded vector subspace and choose an adapted basis $\{u_i,w_a,v_a\}$ such that $l_1 v_a=w_a$ and $u_i$ form a basis in $\imath \bar{L}\subset L$. It follows, $\imath \bar{L}$ is determined by linearly independent linear homogeneous equations $\{w^\a=0,v^\a=0\}$, where $u^i,w^a,v^a$ are coordinate functions dual to basis elements
$\{u_i,w_a,v_a\}$ and $a$ belongs to a countable ordered set. Not that $l^*_1 (w^a)=v^a$ for all $a$. Now we consider a modified embedding of $\bar L$ into $L$ determined by $\{w^a=0, Q_L w^a=0\}$. Using the filtration of the $Q$-complex of functions $\cF(L[1])$ by the polynomial powers one can show that $\{w^a, Q_L w^a\}$ are again functionally independent and thus the subset $\{w^a= Q_L w^a=0\}$ is a $Q$-submanifold of $L[1]$, which is isomorphic to $\bar{L}$. The induced $Q$-structure makes $\bar{L}$ into an $L_\infty$-algebra (the minimal model of $L$)
and the new inclusion $\bar{L}\hookrightarrow L$ into a quasi-isomorphism of $L_\infty$-algebras. Such a procedure is known in mathematical literature as the homotopy transfer.

\begin{example}\label{ex:simple_Koszul} Let $\chi: X\to \Sigma$ be a vector bundle. Consider the pull-back bundle $V\colon =\chi^* (X)=X\times_\Sigma X$
over the total space $X$; $V$ admits the canonical section $F$, induced by the diagonal embedding $X\hookrightarrow X\times_\Sigma X$,
the zero locus of which, $\Sigma=F^{-1}(0)$, coincides with the zero section of $\chi$.
One can easily verify that $\left(\Lambda^\bullet {V^*}, \de=\iota_F\right)$ is the Koszul resolution of $\Sigma$, i.e.
\beq
H_\de^i (\Gamma (\Lambda^\bullet {V^*}))=
\left\{
\be{cc}
C^\infty (\Sigma)\, , & i=0 \\
0\,, i\ne 0
\ee
\right.
\eeq
 If we impose that sections of $\Lambda^i V^*$ have the degree $-i$, so that the whole space of sections
$\Gamma\left(\Lambda^\bullet V^*\otimes \Lambda^\bullet\g^*\right)$ becomes isomorphic to the algebra of functions on
$M=V[-1]$, then $(M, \de)$ is a 
non-positively graded $Q$-manifold. Furthermore, using the projection $M\to\Sigma$ and the embedding $\Sigma\to M$,
uniquely determined by the vector bundle structure $\chi$, one can check that $(M,\de)$ and $(\Sigma, 0)$ are equivalent $Q$-manifolds.

More precisely, let $U$ be an open subset of $\Sigma$ with local coordinates $z^a$ and $q^\mu$ be some linear fiber coordinates on $X|_U$.
The associated local coordinates on $V$ are $(z^a, q^\mu, (q')^\nu)$, such that the canonical section $F$ is given by
$(z^a,q^\mu)\mto (z^a, q^\mu, q^\mu)$. Let $p^\mu$ be local fiber coordinates on $V[-1]$ corresponding to $(q')^\mu$,
so that the degree of $p^\mu$ is equal to $-1$. Then
$\de$ will take the form $\de =\sum_\mu q^\mu \partial_{p^\mu}$.
\end{example}

\begin{example}\label{ex:simple_Koszul2}  Let $\g$ be a Lie algebra, $X\to \Sigma$ be a $\g-$equivariant bundle. Then $F$ from Example \ref{ex:simple_Koszul} is an equivariant section of
a $\g-$equivariant vector bundle $V\to X$.
Let $\left(\Gamma\left(\Lambda^\bullet V^*\otimes \Lambda^\bullet\g^*\right), \gamma\right) $ be the Chevalley--Eilenberg complex, corresponding
to the $\g-$action on sections of $\Lambda^\bullet V^*$ and $s=\de+\gamma$, where the Koszul operator $\de$ is extended to the
whole space by linearity. Then $M=V[-1]_X\times \left(\g[1]\times X\right)$  is a 
$Q$-manifold, which is equivalent to $(\g [1]\times \Sigma, \gamma_0)$.
Here $\gamma_0$ is the Chevalley--Eilenberg differential, corresponding to the $\g$-action on $\Sigma$.
\end{example}

\begin{deff} \label{def:offshell_gauge}  Let $(M,Q)$ be a $Q$-manifold, $\xi$ be a degree $-1$ vector field on $M$.
 An infinitesimal gauge symmetry generated by $\xi$ is the degree zero vector field $\de_\xi =[\xi, Q]=\xi Q+Q\xi$ (cf.~\cite{Bojowald:2004wu} and also~\cite{Grigoriev:1998gn} in the Lagrangian case.).
\end{deff}

Let $\iota\colon (N, Q_N)\hookrightarrow (M,Q)$ be an embedded $Q$-submani\-fold, i.e. $N\hookrightarrow M$ is an embedded $\Z$-graded submanifold
such that $Q|_N=Q_N$. Let us consider $T_N= \left(TM\right)|_N$ as a graded vector bundle over $N$. A section of degree $k$ of $T_N$ can be viewed as
a $\iota$-derivation of $\cF(M)$ with values in $\cF(N)$, that is, a degree $k$ linear operator $v\colon \cF(M)\to \cF(N)$, which satisfies the $\iota$-Leibniz rule
$v (fh)=v(f)\iota^* (g) +(-1)^{k\deg f}\iota^* (f)v(h)$ for any two functions $f,h$ on $M$, where the first function is of pure degree.
Given a vector field $\eta$ on $M$, its restriction onto $N$ is a section of $T_N$, corresponding to the $\iota$-derivation $\iota^*\circ \eta \colon \cF(M)\to \cF(N)$.
The linearization of $Q$ at $N$ defines a nilpotent degree $1$ bundle map $T_N\to T_N$, $v\mto v\circ Q-(-1)^k Q_N \circ v$ for any $v\in\Gamma(T_N)^k$.

\begin{deff} \label{def:onshell_gauge}  Let $\e$ be a degree $-1$ section of $T_N$.
An infinitesimal gauge symmetry at $N$ generated by $\e$ is the degree zero
$\iota$-derivation $\de_\e =\e\circ Q + Q_N \circ \e$.
\end{deff}

The nilpotency condition for $Q$ asserts that any infinitesimal gauge symmetry commutes with $Q$,
therefore the corresponding infinitesimal flow preserves the subspace of $Q$-submanifolds. The proof of the next statement is straightforward.

\begin{proposition}\label{prop:offshell_vs_onshell}
Given a $Q$-submanifold $N$ of $M$, the restriction of any gauge symmetry  $\de_\xi$ onto $N$
is an infinitesimal gauge symmetry at $N$, generated by $\e= \de_\xi|_N$.
\end{proposition}

\subsection{PDEs}

By definition PDE is a pair $(E_X,\cC)$, where $E_X$ is a manifold and $\cC(E_X)$ (denoted by just $\cC$ in what follows if it doesn't lead to confusions) is an involutive distribution $\cC(E_X)\subset TE_X$ called Cartan distribution. It is typically assumed (as it's done later) that
\smallskip
\begin{enumerate}[i)]
 \item $E_X$ is a locally trivial bundle $\pi_X: E_X \to X$ over the manifold $X$ of independent variables.
 \item Canonical projection $\pi_X$ induces an isomorphism  $\cC_p(E_X) \to T_{\pi_X(p)}X$ for all $p\in E_X$. In particular $\cC$ is of constant rank, which is equal to $\dim(X)$.
 \item $(E_X,\cC)$ can be embedded into some jet bundle as an infinitely prolonged equation, at least locally.
\end{enumerate}
\smallskip
For an infinitely prolonged PDE realized as a submanifold of the respective jet-bundle (in particular, the jet-bundle itself) the Cartan distribution is canonical. For a modern review see e.g.~\cite{Krasilshchik:2010sst}.

It is useful to consider the algebra $\Omega_h(E_X)$ of horizontal differential forms on $E_X$, i.e. forms on $E_X$ that vanish on vertical vectors. This can be seen as generated by functions on $E_X$ and differential forms on $X$ pulled back by the canonical projection. It is also convenient to think of $\Omega_h(E_X)$ as functions on the supermanifold $E=\cC[1](E_X)$, i.e. the total space of the Cartan distribution with the reversed parity of fibers. If $x^a$ are local coordinates on $X$ then $\pi_X^*({\rm d}x^a)$ are basis horizontal forms which we denote just by ${\rm d}x^a$ in what follows.

The total derivatives along $x^a$, denoted by $D_a$,
are (locally defined) vector fields generating $\cC(E_X)$, such that $(\pi_X)_* (D_a)=\dl{x^a}$. For instance, if $E_X$ is the jet-bundle of the trivial vector bundle $X\times \fR^1 \to X$, then
\begin{equation}
 D_a=\dl{x^a}+\phi_a\dl{\phi}+\phi_{ab}\dl{\phi_b}+\cdots
\end{equation}
where $\phi$ is the fiber coordinate and $\phi_{a\cdots}$ are its `derivatives'. In terms of local coordinates the horizontal differential reads as
\begin{equation}
 \dh={\rm d}x^a D_a
\end{equation}
and can be considered as a $Q$-structure on $E$. 

Homological vector field $\dh$ encodes the Cartan distribution so that in the language of supermanifolds PDE can be defined as $(E,\dh)$. Moreover, $E$ can be thought of as a super-bundle over $T[1]X$ in which case the canonical projection is a morphism of $Q$-manifolds: $(E,\dh)\to (T[1]X,{\rm d}_X)$. This condition precisely implies that the projection induces the isomorphism $\cC_p(E_X)\to T_{\pi_X(p)}X$.

Vector fields on $E_X$ belonging to the Cartan distribution can be represented as restrictions to $E_X$ of vector fields on $E$ of the form:
\begin{equation}
 \commut{{\rm d}_h}{Z}\,, \qquad {\rm deg}(Z)=-1
\end{equation}
Evolutionary vector fields (i.e. preserving Cartan distribution) on $E_X$  are then in one-to-one correspondence with $\dh$-cohomology $H^0(\commut{\dh}{\cdot})$ of the space of vector fields on $E$. In other words symmetries of the PDE $(E,\dh)$ are
precisely the above cohomology. Note that each cohomology class has a representative which is vertical. Indeed, a horizontal piece $H^a D_a$ of $Z$ can be removed by subtracting $\big[\dh,H^a\dl{({\rm d}x^a)}\big]$.

\subsubsection{Intrinsic representation of PDE}
\label{sec:intrinsic}

Given PDE $(E_X,\cC)$ defined in terms of intrinsic geometry of $E_X$ (i.e. without referring to one or another embedding of $E_X$ into a jet-bundle) one can construct a jet-bundle in terms of $E_X$ and embed $(E_X,\cC)$ as an infinitely-prolonged equation therein. More precisely, consider $J^1(E_X)$
(recall that $E_X$ is a bundle over $X$). The Cartan distribution on $E_X$ can be represented as a $1$-form on $E_X$ with values in vertical tangent vectors such that it is zero on $\cC_p(E_X)$ and acts identically on vertical vectors. This form can be regarded as a connection 1-form $\Gamma$ of the Cartan distribution seen as an Ehresmann connection. The involutivity of the Cartan distribution is equivalent to the flatness of the connection.

At the same time $\Gamma$ can be viewed as a section of $J(E_X)\to J^1(E_X)$. The image of this section is a submanifold $\cE\subset J^1(E_0)$ which is by construction diffeomorphic to $E_X$. Being a submanifold of the jet-bundle,  $\cE$  defines a new equation. It is easy to check that this equation is just a new form of $(E_X,\cC)$, i.e. is equivalent to $(E_X,\cC)$.

Let us write down explicitly the new form of the equation. If $\psi^A$ are coordinates on the fibers of $E_X$, vector fields $D_a$ can be locally written as
\begin{equation}
D_a=\dl{x^a}+\Gamma_a^A(x,\psi)\dl{\psi^A}
\end{equation}
for some functions $\Gamma_a^A(x,\psi)$. Of course $\Gamma_a^A(x,\psi)$ are just nontrivial components of the connection 1-form.  The remaining components $\Gamma_A^B=\delta_A^B$ thanks to the condition that it acts identically on vertical vectors. The submanifold $\cE\subset J^1(E_X)$ is determined by the constraints
\begin{equation}
 \psi^A_a=\Gamma_a^A(x,\psi)\,.
\end{equation}

Switching to the standard language of dependent and independent variables this PDE has $x^a$ as independent variables, $\psi^A$ as dependent and the equations read explicitly:
\begin{equation}
\label{cov-const}
\dl{x^a}\psi^A(x)=\Gamma_a^A(x,\psi(x))
\end{equation}
It is easy to check that equation is equivalent to the starting point $(E_X,\cC)$. This form of the equation can be regarded as the covariant constancy form.

Let us give a supergeometrical interpretation of the intrinsic representation. To this end let us consider $E_{T[1]X}=\cC[1](E_X)$ as a superbundle over $T[1]X$. Let $\sigma$ be a section of this bundle which preserves the degree, i.e. $\sigma^*$ preserves the degree. The condition that $\sigma$ is a solution has a simple geometrical meaning:
\begin{equation}
\sigma^*\circ \dh={\rm d}_X \circ \sigma^*
\end{equation}
where ${\rm d}_X$ is the canonical $Q$-structure on $T[1]X$ (de Rham differential on $X$). In other words $\sigma$ is a $Q$-map. If $\psi^A$ are local coordinates on the fibres then applying the above equality to $\psi^A$ one again gets:
\begin{equation}
 \dl{x^a}\psi^A(x)=(\dh\psi^A)|_{\psi^B=\psi^B(x)}=\Gamma_a^A(x,\psi(x))\,,
\end{equation}
where $\psi^A(x)=\sigma^*(\psi^A)$. This is precisely the covariant constancy equation~\eqref{cov-const}.

\subsection{Standard gauge PDEs}
\label{sec:sgPDE}

To motivate the introduction of gauge PDE as a geometrical object let us recall what is typically called classical local gauge field theory in physics literature. Instead of starting with equations of motion, gauge symmetries, and gauge for gauge symmetries and then constructing BV formulation we immediately start with BV. More precisely, consider the space of fields, ghosts, antifields, etc., which in geometrical terms is a graded locally trivial bundle $F_X\to X$ over space-time manifold $X$.  In addition $F_X$ is assumed to be finite-dimensional though some reasonable generalization such as locally finite-dimensional bundles can be also allowed.  This data gives rise to the jet-bundle $E_X=J^\infty(F_X)$ over $X$ which is also a graded and locally trivial bundle over $X$, the grading is called ghost degree. All the information about the theory is contained in the BV-BRST differential $s$ which is assumed  nilpotent, vertical, evolutionary and of ghost degree $1$. In what follows we refer to this geometrical data as to standard gauge pre-PDE. If the theory is Lagrangian, in addition one requires $E_X$ to be equipped with an odd Poisson bracket of ghost degree $1$ and $s$ to be Hamiltonian, giving the usual Batalin--Vilkovisky formulation of the system.

It is instructive to see how equations of motion and gauge symmetries are encoded in the homological vector field $s$. To this end let us introduce local coordinates $x^a,\phi^i,c^\alpha,\cP^\mu,\ldots$ on the underlying bundle $F_X$ (seen as 0-th jets) such that $x^a$ are coordinates on the base, $\phi^i$ coordinates of degree zero (fields), $c^\alpha$ of degree $1$ (ghosts), $\cP^\mu$ of degree $-1$ (antifields). Note that in general there can be coordinates of higher and lower ghost degrees, which are responsible for relations between the equations, gauge transformations and their higher analogs. Then the equations of motion and gauge symmetries can be explicitly written as~\cite{Barnich:2004cr}:
\begin{equation}
\label{eq-gs-coord}
 (s\cP^\mu)|_{\Psi^k=0\,, k\neq 0}=0\,, \quad \delta_\epsilon \phi^i=(s\phi^i)|_{\Psi^k=0\,, k\neq 0,1\,, c^\alpha\to \epsilon^\alpha}\,,
\end{equation}
where $\Psi^k$ denote all the coordinates of ghost degree $k$ and in the second formula ghost fields $c^\alpha$ and their derivatives are to be replaced by gauge parameters $\epsilon^\alpha(x)$ and their derivatives. In a similar way one defines higher order (gauge for gauge) gauge transformations.

In more geometrical terms solutions are parallel (with respect to Cartan distribution) degree zero sections of $E_X \to X$ such that $s$ vanishes on them. It is also useful to define the stationary surface, which is the body (i.e. degree zero submanifold) of the zero locus of $s$. In these terms solutions are precisely those sections whose prolongations belong to the stationary surface.

To get a more geometrical understanding of the gauge transformations let $\xi$ be a vertical evolutionary vector field of ghost degree $-1$. Such field is always a prolongation of a vertical field $\xi_0$ on $F_X$ which serves as a gauge parameter in this setting. Just like in the case of $Q$-manifolds the infinitesimal gauge transformation associated to $\xi$ is an evolutionary vector field $\commut{s}{\xi}$. This vector field clearly restricts to the body of $E_X$. Indeed, the body of $E_X$ is locally determined by equations $\Psi^{k\neq 0}=0$.
Because $\commut{s}{\xi}$  carries vanishing degree, $\commut{s}{\xi}\Psi^{k}$ carries nonvanishing degree for $k\neq 0$ and hence vanishes on the body of $E_X$. It is important to distinguish infinitesimal gauge transformations understood as vector fields on the entire $Q$-manifold and those on its body or the stationary surface.

To see the relation with~\eqref{eq-gs-coord} let us take $\xi_0=\epsilon^\alpha(x)\dl{c^\alpha}$. It is easy to check that the restriction of $\commut{s}{\xi}$
to the body of $E_X$ coincides with the second formula in~\eqref{eq-gs-coord}:
\begin{equation}
\begin{aligned}
(\commut{s}{\xi}\phi^i)\big|_{\Psi^k=0\, k\neq 0}&=({\xi}{s}\phi^i)\big|_{\Psi^k=0\, k\neq 0}\\
&=({s}\phi^i)\big|_{\Psi^k=0\, k\neq 0,1\,, c^\alpha\to \epsilon^\alpha}\,.
\end{aligned}
\end{equation}

Just like in the case of usual PDEs it is useful to switch to the language of $Q$-bundles. To this end let us extend  $J^\infty(F_X)$ to the bundle $E_{T[1]X}$ over $T[1]X$. Now the total space is equipped with a bidegree $(p,k)$, where $p$ is the form degree (i.e. homogeneity in $\theta^a\equiv {\rm d}x^a$) and $k$ is the original ghost grading. The condition that section $\sigma: T[1]X \to E_X$ is a solution is equivalent to that $\sigma$ is a $Q$-map $(T[1]X,{\rm d}_X)\to (E_{T[1]X},\dh+s)$ of bidegree $(0,0)$. Indeed, vanishing bidegree implies that $\sigma$ is a $Q$ map for $\dh$ and $s$ separately. 

An important observation is that in this setting it is sufficient to take care of the total degree only.  More precisely, let us consider a $Q$-section $\sigma$ of total degree $0$ and show that it is gauge equivalent (with the parameter of total degree $-1$) to a bidegree $(0,0)$ section.  The natural gauge equivalence of such sections is defined as follows: let $\chi^*$ be a degree $-1$ map $\cF(E_{T[1]X})\to \cF(E_{T[1]X})$ satisfying $\chi^*(fg)=\chi^*(f)\sigma^*(g)+(-1)^{\p{f}}\sigma^*(f)\chi^*(g)$ and $\chi^*(x^a)=\chi^*(\theta^a)=0$. It plays a role of gauge parameter. The infinitesimal gauge transformation then reads as
\begin{equation}
 \delta \sigma^*=\md_X\circ \chi^*+\chi^*\circ Q\,.
\end{equation} 
It is easy to see that this preserves $Q$-map condition.

To analyze explicitly the equations and gauge equivalence
it is convenient to extend $F_X$ to a bundle $F_{T[1]X}$ over $T[1]X$ and to work with super jet-bundle $J^\infty(F_{T[1]X})$. It is equipped with the total ghost degree and the horizontal degree arising from the standard grading on $T[1]X$. Sections of $E_{T[1]X}$ of total degree zero can be identified with bidegree $(0,0)$ sections of 
$J^\infty(F_{T[1]X})$. 

Let us introduce  coordinates $x^a,\theta^a,\psi^A_{(a_1\cdots a_l)|[b_1\cdots b_k]}$ on $J^{\infty}(F_{T[1]X})$, 
Note that the total degree of $\psi^A_{(a_1\cdots a_l)|[b_l\cdots b_k]}$ is $\gh{\psi^A}-k$. It is convenient to pack $\psi$-coordinates into the following generating functions:
\begin{multline}
\label{decomposition}
\Psi^A(y,\xi)=\sum_{k,l\geq 0} \frac{1}{k!l!}
\xi^{a_k}\cdots\xi^{a_1}y^{b_l}\cdots y^{b_1}\psi^A_{(b_1\cdots b_l)|[a_1\cdots a_k]}\\
\equiv  \xi^{(\nu)} y^{(a)}\psi^A_{(b)|[\nu]}\,.
\end{multline}
In the above local coordinates a section of $E_{T[1]X}$ is locally a collection of functions $\psi^A_{(a_1\cdots a_l)|[b_l\cdots b_k]}(x)$ and it is convenient to parameterize it in terms of $\Psi^A(x,y,\xi)$. 

In these terms it is useful to introduce the following locally defined (differential) operators on sections:
\begin{equation}
 {\rm d}^F\Psi=\Psi \left(\dr{x^a}\xi^a \right)\,, \qquad \sigma^F\Psi=\Psi \left(\dr{y^a} \xi^a\right)\,,
\end{equation}
For instance:
\begin{equation}
 {\rm d}^F\psi^A_{()|[b]}=D_b\psi^A_{()|[]}\,,\qquad  \sigma^F\psi^A_{()|[b]}=\psi^A_{(b)|[]}\,.
\end{equation}

If we denote by $\psi^{\alpha_{k}}_k$ all the $\psi^A_{(a)|[b]}$ of total degree $k$, the condition that  $\psi^{\alpha_0}_0(x)$ determine a $Q$-section takes the form:
\begin{equation}
\label{eq-p}
 \left(({\rm d}^F-\sigma^F+\bar s)\psi^{\alpha_{-1}}_{-1}\right)\big|_{\psi^{\alpha_k}_{k}=0\,, k\neq 0}=0\,,
\end{equation}
where $\bar s$ denote a  natural prolongation of $s$ from $J^{\infty}(F_{X})$ to $J^{\infty}(F_{T[1]X})$.
At the same time infinitesimal gauge transformation reads as
\begin{equation}
 \delta \psi^{\alpha_{0}}_0=\left(({\rm d}^F-\sigma^F+\bar s)\psi^{\alpha_{0}}_{0}\right)\big|_{\psi^{\alpha_k}_{k}=0\,, k\neq 0,1\,, \psi^{\alpha_1}_{1}=\epsilon^{\alpha_1}(x)}\,.
\end{equation}

All the coordinates $\psi^A_{(a_1\cdots a_l)|[b_l\cdots b_k]}$ besides $\psi^A_{()|[]}$ can be split into two subsets $w^q,v^q$ such that
$\sigma^F w^q=v^q$. It is also convenient to denote by $w^{q_k}_k$ and $v^{q_k}_k$ those of degree $k$. Consider the following subset of equation~\eqref{eq-p}:
\begin{equation}
 \left(({\rm d}^F-\sigma^F+ \bar s)w^{q_{-1}}_{-1}\right)|_{\psi^{\alpha_k}_{k}=0\,, k\neq 0}=0\,.
\end{equation}
This determines $v^{q_0}_0$ in terms of other variables. Furthermore, analyzing gauge transformations one finds that
\begin{equation}
 \delta w^{q_0}_0=-\epsilon^{a_1}+\cdots\,,
\end{equation}
which upon using suitable degree implies that $w^{q_0}_0$ can be set to zero. Again using a suitable degree one can analyze the remaining equations order by order and show that after setting $w^{q_0}_0=0$ all $v^{q_0}_0$ that carry positive form degree also vanish (the form degree is introduced according to $\deg\big({\psi^A_{(a_1\cdots a_l)|[b_1\cdots b_k]}}\big)=k$.

This shows that any $Q$-section preserving the total degree is equivalent to a section preserving bidegree using the equivalence relation. What we just demonstrated is that the gauge theory defined in terms of $(E_X,s)$ is equivalent to the one whose fields are sections preserving the total degree and equations of motion are conditions that the section is a $Q$-morphism. A remarkable fact is that this equivalent theory also admits a concise BV formulation. Indeed, one simply takes $s^P={\rm d}^F-\sigma^F+\bar s$ (extended to a vector filed on super-jet bundle $J^{\infty}(E_{T[1]X})$) to be its BV-BRST differential and $J^{\infty}(E_{T[1]X})$ as a jet-bundle. This is again
a standard gauge PDE, known as the parent formulation~\cite{Barnich:2010sw}, which is equivalent to the original gauge theory. This equivalence as well as the above considerations explicitly made use of coordinates and hence work only locally. One of the goals of the present work is to propose a globally well-defined notion of a gauge PDE and parent formulation.

To summarize, given a standard gauge PDE one can either use the standard interpretation where solutions are $Q$-sections preserving  bidegree or define solutions to be $Q$-sections preserving only the total degree. In so doing one should also adjust accordingly  (higher) gauge transformations. The two interpretations are equivalent. The second of them has an advantage because it does not require extra degree and hence is more flexible. In the next section we develop an approach to gauge PDEs using this more flexible interpretation.

\begin{example}[Standard form of Maxwell gauge PDE]\label{ex:maxwell}
To illustrate the notion of gauge PDE consider a simple example of Maxwell equations seen as a gauge theory. The geometrical data determining the system is a pseudo-Riemannian manifold $X$ which we take to be Minkowski space for simplicity. The fiber bundle $F_X$ is a direct sum of $TX\oplus U$, where $U$ is a trivial vector bundle with $1$-dim fiber of degree $1$, and its dual vector bundle with the degree shifted by $-1$. As local coordinates on the fibers we take $A_b,C,\cP,A_*^b$ of the following ghost degree:
\begin{equation}
\begin{aligned}
 \gh{A_b}&=0\,,\\ 
 \gh{C}&=1, \\ 
 \gh{\cP}&=-2 \,,\\ 
 \gh{A_*^b}&=-1\,.
\end{aligned}
\end{equation}
The BV-BRST differential is an evolutionary vertical vector field on $J^\infty(F_X)$ determined by
\begin{equation}
\begin{aligned}
 sA_*^b&=D_c (D^c A^b-D^b A^c)\,,\\ 
 s A_a&=D_a C\,, \\
 s \cP&=D_c A^c_*
\end{aligned}
 \end{equation}
where $D_a$ denotes the total derivative.  Let $\sigma$ be a section $X\to J^\infty(F_X)$ of vanishing degree.
The condition that $\sigma$ is a solution says that $\sigma$ is parallel and $s$ vanish at its image. The first condition implies that $\sigma$ is a prolongation of a section $\sigma_0:X\to F_X$:
\begin{equation}
\begin{aligned}
 \sigma_0^*(A_c)&=A_c(x)\,,\\
 \sigma_0^*(C)&=0\,, \\ 
 \sigma_0^*(\cP)&=0\,, \\
 \sigma_0^*(A^a_*)&=0\,.
\end{aligned}
\end{equation}
Asking that $s$ vanishes on the image of $\sigma$ amounts to $\d_a(\d^b A^c(x)-\d^c A^b(x))=0$, i.e. Maxwell equation.  Replacing (total derivatives of) $C$ in $sA_a$ with (derivatives of) gauge parameter $\epsilon(x)$ according to~\eqref{eq-gs-coord} one arrives at the standard gauge transformation law: $\delta_\epsilon A_a(x)=\dl{x^a}\epsilon(x)$.
\end{example}

\section{Gauge PDE as a $Q$-bundle}
\label{sec:Q-bundle}

Now we are ready to introduce the notion of a gauge PDE, which is more flexible than standard gauge PDEs discussed in the previous Section. We first introduce more general objects, gauge pre-PDEs, and then define their equivalences. Then we define gauge PDEs as those gauge pre-PDEs which satisfy some extra conditions formulated using the equivalence.
\begin{deff}
Gauge pre-PDE is a $\fZ$-graded $Q$-bundle $(E_{T[1]X},Q)$ over $(T[1]X,{\rm d}_X)$, where $(T[1]X,{\rm d}_X)$ is considered as a graded $Q$-manifold with the canonical degree (form degree) and the canonical $Q$-structure (de Rham differential).
\end{deff}
For simplicity in what follows we restrict to the case where $(E_{T[1]X},Q)$ is a locally trivial $\fZ$-graded bundle over $T[1]X$ (but we do not require it to be locally trivial as a $Q$-bundle).

To this end we need the following:
\begin{deff}
Gauge pre-PDE $(E_{T[1]X},Q)$ is called contractible if as a bundle over $T[1]X$ it is locally trivial, admits a global $Q$-section, and its fiber is a contractible $Q$-manifold.
\end{deff}
Let us explicitly write down the local structure of a contractible PDE in terms of local coordinates. By definition one can find local coordinates (local trivialization) $x^a,\theta^a,w^I,v^I$ such that
\begin{equation}
 Qx^a=\theta^a\,,\qquad Qw^I=v^I\,.
\end{equation}

Now the equivalence is defined as
\begin{deff}\label{def:equivalence_pre-gauge}
Gauge pre-PDE $(E_{T[1]X},Q)$ is an equivalent reduction of $(E^\prime_{T[1]X},Q^\prime)$ if
$(E^\prime_{T[1]X},Q^\prime)$ is a locally-trivial $Q$-bundle over $(E_{T[1]X},Q)$ (in the category of $Q$-bundles over $T[1]X$)
whose fiber is contractible and which admits a global $Q$-section $i:E_{T[1]X}\to E^\prime_{T[1]X}$. The
equivalence relation generated by the equivalence reduction is called the equivalence of gauge pre-PDEs.
\end{deff}

\begin{proposition} Two gauge pre-PDEs are equivalent if and only if there exists a third one such that the two are its equivalent reductions.
\end{proposition}

\noindent\proof
It is sufficient to check the transitivity property, i.e. to show that, if $(E_1,Q_1)$ and $(E_2, Q_2)$
are equivalent reductions of $(E',Q')$, $(E_2,Q_2)$ and $(E_3, Q_3)$ are equivalent reductions of $(E'',Q'')$\footnote{For simplicity we omit the
subscript $T[1]X$ as it is clear that all these bundles are defined over $T[1]X$.}, then there exists a gauge pre-PDE $(E,Q)$ such that
 $(E_1,Q_1)$ and $(E_3, Q_3)$ are equivalent reductions of $(E,Q)$.

First, let us notice that the composition of two subsequent equivalent reductions is an equivalent reduction. On the other hand, the fibered product of two $Q$-bundles over an arbitrary $Q$-manifold is again a $Q$-bundle over the same base. By use of the latter,
we take $(E,Q)\colon =(E',Q')\times_{(E_2,Q_2)} (E'',Q'')$.

Indeed, it is easy to verify that
the canonical projections $(E,Q)\to (E',Q')\to (E_1,Q_1)$ and $(E,Q)\to (E'',Q'')\to (E_3,Q_3)$ have contractible fibers. Furthermore,
$(E,Q)\to (E_1,Q_1)$ admits the canonical $Q$-section given by the composition of sections $(E,Q_1)\to (E',Q')$ and $(E',Q')\to (E,Q)$,
where the last one is induced by the $Q$-section $(E_2,Q_2)\to (E'',Q'')$ and the $Q$-morphism $E'\to E'\times E''$. Similarly,
$(E,Q)\to (E_1,Q_1)$ admits the canonical $Q$-section which is defined in like manner. Thus  $(E_1,Q_1)$ and $(E_3, Q_3)$
are equivalent reductions of $(E,Q)$. This finally proves the proposition.
\hfill $\square$

\smallskip

In particular, a contractible gauge pre-PDE is always equivalent to empty pre-PDE $T[1]X$. In agreement with Section~\bref{sec:sgPDE} we call a pre-PDE $(E_{T[1]X},Q)$ standard if $E_{T[1]X}$ is an extension  to $T[1]X$
of a bundle $J^\infty(F_X) \to X$, with $F_X$ (locally) finite-dimensional, and $Q=\dh+s$ where $\dh$ is a canonical horizontal differential on the jet-bundle and $s$ is vertical. 

\begin{deff}
Gauge pre-PDE $(E_{T[1]X},Q)$ is a gauge PDE if it is equivalent to a: 
\smallskip
\begin{enumerate}[i)]
\item nonnegatively graded gauge pre-PDE and
\item standard gauge pre-PDE
\end{enumerate}
\end{deff}
Let us see what a gauge PDE looks like in terms of local trivialization. By assumption $E_{T[1]X}$ as a graded manifold can be represented locally as a product of $T[1]X$ (with canonical grading) and a typical fiber $H$ in such a way that the $\fZ$-degree is the total degree. In other words local trivialization induces a local noncanonical bidegree. According to this bidegree $Q$ decomposes as follows
$
Q=Q_{0,1}+Q_{1,0}+Q_{2,-1}+\cdots
$
where the first subscript denotes the degree on $T[1]X$ (i.e. form degree). Note that there are no components of negative form degree because these can only originate from terms involving $\dl{\theta^a}$ but these are forbidden because $Q$ projects to ${\rm d}_X$ by the canonical projection. Note that if $Q_{i,j}=0$ for $i>1$ we are back to the conventional situation where $Q_{1,0}$ is interpreted as $\dh$ while $Q_{0,1}$ as $s$. Note that in general $Q_{i,j}\neq 0$ for $i>1$ and moreover the split of $Q$ depend on the choice of local trivialization.

One can consider a restricted class of gauge (pre)-PDE where $E_{{T[1]X}}$ is equipped with bidegree, i.e. the total degree canonically splits into the sum of horizontal degree which projects to form degree on $T[1]X$ and the vertical degree. In this case the decomposition of $Q$ into homogeneous pieces is canonical so that requiring $Q_{i,j}=0$ for $i>1$ one arrives at a particular class which can be called bigraded gauge (pre)-PDE. If in addition $E_{{T[1]X}}$ is a jet-bundle we are back to standard gauge (pre)PDEs.

As we are going to see any gauge PDE can be equivalently represented as a standard gauge (pre)PDE. However, the formalism where the bidegree is not preserved is very convenient.

\begin{example}[Gauge ODE] \label{ex:ODE}
As a toy model example illustrating the flexibility of the formalism let us consider a gauge PDE over $T[1]X$
with ${\rm dim}(X)=1$. Let us restrict ourselves to the case where $E_X$ is finite-dimensional (which is not a severe restriction in $1d$) and introduce local coordinates $x,\theta,\psi^\alpha$, where $x,\theta$ are pullbacks of standard coordinates on $T[1]X$. The general form of $Q$ which projects to ${\rm d}_X=\theta\dl{x}$ on $T[1]X$ reads as
\begin{equation}
\begin{aligned}
Q&=\theta\dl{x}+\theta V+ q\,, \\ 
V&=V^\alpha (x,\psi)\dl{\psi^\alpha}\,, \\
q&=q^\alpha(x,\psi)\dl{\psi^\alpha}\,.
 \end{aligned}
\end{equation}
The terms $\theta V$ and $q$ have a clear interpretation of  the `evolution' vector field and the BRST differential $q$ respectively. In one or another version this form showed up in the literature~\cite{Barnich:2010sw,Kaparulin:2012ie}. In particular, its counterpart in the case of Hamiltonian/Lagrangian case was already in~\cite{Grigoriev:1999qz}.
\end{example}

\begin{example}[Minimal form of Maxwell gauge PDE] In the setting of example~\bref{ex:maxwell} let us describe the same gauge PDE as a $Q$-bundle which is not any longer a jet-bundle and where the bidegree is not preserved. As before let $x^a,\theta^a$ denote local coordinates on $T[1]X$ and let the fiber be a superspace with coordinates $C,F_{ab},F_{ab;c},\ldots F_{ab;c\cdots}$ such that 
$\gh{C}=1$, $\gh{F}=0$, $F_{ab;c_1\cdots c_k}=F_{a(b;c_1\cdots c_k)}$, with $()$ denoting symmetrization, and all $F_{\ldots}$ are totally traceless with respect to the Minkowski space metric. The $Q$-structure on the total space $E_{T[1]X}$ is determined by
\begin{equation}
\label{Q-Maxwell}
\begin{aligned}
Qx^a&=\theta^a\,,\\
QC&=\half \theta^a\theta^b F_{ab}\,, \\ 
QF_{ab}&=\theta^c F_{ab,c},\\
&\kern4pt\vdots
\end{aligned}
\end{equation}
The condition that $\sigma:T[1]X\to E_{T[1]X}$ is a $Q$-map takes the form
\begin{equation}
\begin{aligned}
 {\rm d}_X A(x,\theta)&=\tfrac12\theta^a\theta^b F_{ab}(x,\theta)\,, 
 \\
 {\rm d}_XF_{ab}(x)&=\theta^c F_{ab,c}(x)\,,
 \\
&\kern4pt\vdots
\end{aligned}
\end{equation}
where $A(x,\theta)$ and $F_{ab,\cdots}(x)$ are introduced as $\sigma^* C=A(x,\theta)=A_a\theta^a$ and $F_{ab,\cdots}(x)=\sigma^*(F_{ab,\cdots})$.
Taking the trace of the second equation and using the first one, one immediately arrives at the Maxwell equation on $A_a(x)$.

The $Q$-manifold (BRST complex) determined by $Q$ on $E_X$ and its generalization to YM theory and Einstein gravity  has been actively discussed in \cite{Brandt:1996mh,Brandt:2001tg} while the first relation in \eqref{Q-Maxwell} is the version of so-called `Russian formula' \cite{Stora:1983ct}. This formulation is also closely related to the unfolded formulation~\cite{Vasiliev:1988xc,Vasiliev:2005zu}. Note that \eqref{Q-Maxwell} is a minimal BRST complex in the sense that one cannot reduce it further (at least in the space of local functions).
\end{example}

\subsection{AKSZ-type sigma models}

An interesting class of gauge PDEs is provided by so-called AKSZ-type sigma models. Originally the term AKSZ sigma model refers to Lagrangian topological gauge theories of certain structure and finite number of fields~\cite{Alexandrov:1995kv}. Now, following~\cite{Barnich:2005ru,Barnich:2010sw},  we use it to refer to gauge (pre)PDEs of special form. More specifically, the data of AKSZ-type sigma model is given by a trivial $Q$-bundle, i.e.
$(E_{T[1]X},Q)=(T[1]X,{\rm d}_X)\times (M,Q_0)$, where $M$ is the fiber. In the sigma-model terminology the fiber is called the target space while $(T[1]X,{\rm d}_X)$ the source space. Fields of AKSZ-type sigma model are degree zero maps
$\sigma: T[1]X \to M$. The equations of motion are the $Q$-map conditions:
\begin{equation}
\label{AKSZ-eom}
 \sigma^*\circ Q_0={\rm d}_X\circ \sigma^*\,,
\end{equation}
where $\sigma^*$ is a pullback associated to $\sigma$. If $\sigma$ is a fixed map, a gauge parameter determining a gauge transformation of $\sigma$
is a degree $-1$ map $\xi^*:\cF(M)\to \cF(T[1]X)$ that satisfies the relation
\begin{equation}
 \xi^*(fg)=(\xi^*(f))\sigma^*(g)+(-1)^{\p{f}}(\sigma^*(f))\xi^*(g)\,, 
\end{equation}
$\forall f,g\in \cF(M)$. The infinitesimal gauge equivalence transformation of $\sigma$ can be written as:
\begin{equation}
\label{AKSZ-gs}
 \delta \sigma^*= {\rm d}_X\circ \chi^*+\chi^* \circ Q_0\,.
\end{equation}
In a similar way one can define gauge equivalence of gauge parameters and its higher analogs. A natural generalization~\cite{Bonavolonta:2013mza} of AKSZ sigma models is achieved by replacing  the space of maps from $T[1]X$ to $(M,Q_0)$ by the space of sections of a locally trivial $Q$-bundle over $T[1]X$.

An original observation made in~\cite{Alexandrov:1995kv} in the case of  Lagrangian topological theories is that the superspace of maps from $T[1]X$ to $(M,Q_0)$ is equipped with a natural $Q$-structure which turns out to be the  BV-BRST differential encoding the equations of motion~\eqref{AKSZ-eom}, gauge symmetries~\eqref{AKSZ-gs} and (higher) gauge for gauge symmetries. In other words the BV formulation of the underlying gauge theory is immediately arrived at by considering the space of super maps from the source to the target supermanifolds.

More precisely, working in terms of jet-bundles the space of super maps is replaced by super jet-bundle $J^{\infty}(E_{T[1]X})$. In this terms the BV-BRST differential is precisely the vertical part of the prolongation $Q^P$ of the total differential $Q={\rm d}_X+Q_0$ from $E_{T[1]X}$ to $J^{\infty}(E_{T[1]X})$. By considering $J^{\infty}(E_{T[1]X})$ as a bundle over $X$ we arrive at the standard gauge PDE $(J^{\infty}(E_{T[1]X}),Q^P)$ of the special form. This is AKSZ-type sigma model seen as a gauge PDE. In a slightly different terms this realization was discussed in~\cite{Barnich:2009jy,Barnich:2010sw,Bonavolonta:2013mza}.

\begin{example}[Zero curvature equation and Chern-Simons model] Take as $(M,Q_0)$ a Q-manifold $\algg{}[1]$ describing CE complex of a Lie algebra $\algg$. If $c^\alpha$ are coordinates on $\algg[1]$, vector field $Q_0=\half\commut{c}{c}^\alpha\dl{c^\alpha}$. If $x^a,\theta^a$ are local coordinates on $T[1]X$ the degree $0$ map $\sigma:T[1]X \to M$ can be written as $\sigma^*(c^\alpha)=A^\alpha(x,\theta)=\theta^a A^\alpha_a(x)$ and the condition that $\sigma$ is a $Q$-map gives ${\rm d}_X A^\alpha=\half\commut{A}{A}^\alpha$, i.e. the zero-curvature equation. Introducing gauge parameter $\xi$ through
\begin{equation}
 \xi^*(c^\alpha)=\epsilon^\alpha(x)
\end{equation}
and using~\eqref{AKSZ-gs} one arrives at $\delta A^\alpha={\rm d}_X\epsilon^\alpha+\commut{\epsilon}{A}^\alpha$.

Now switch to AKSZ and consider $J^{\infty}(E_{T[1]X})$ as a bundle over $X$. In other words, represent a super map as
\begin{equation}
c^\alpha(x,\theta)= \overset{0}{c}{}^\alpha(x)+ \overset{1}{c}{}_a^\alpha(x)\theta^a+\overset{2}{c}{}_{ab}^\alpha(x)\theta^a\theta^b+\cdots
\end{equation}
A useful coordinate system on the fibers of $J^{\infty}(E_{T[1]X})$ is given by $\overset{k}{c}{}^\alpha_{a_1\cdots a_k}$ and their total $x$-derivatives. The ghost degree is $\gh{\overset{k}{c}{}^\alpha_{a_1\cdots a_k}}=1-k$. In particular, $\overset{1}{c}{}_a^\alpha$ is of vanishing degree and is precisely the component field $A^\alpha$ introduced above. In these coordinates the vertical part $s^P$ of differential $Q^P$ has the structure $s^P={\rm d}^F+\bar Q_0$, where ${\rm d}^F$ is introduced by ${\rm d}^F c^\alpha(\theta)=-{\rm d}_X c^\alpha(\theta)$ (see Appendix~\bref{sec:app-sjets}) and $\bar Q_0$ is the prolongation of $Q_0$. 

In the case where $X$ is 3-dimensional and $\algg$ admits an invariant non-degenerate inner product this becomes a genuine BV description where the space of super maps is odd symplectic and $s^P$ is Hamiltonian.
\end{example}

\subsection{Reparametrization invariant gauge PDEs}

Among gauge PDEs there is a special class for which $(E_{T[1]X},Q)$ is a locally-trivial $Q$-bundle. For instance, any PDE of a finite type satisfying certain regularity conditions corresponds to a locally-trivial $Q$-bundle. 
Let us consider ODE of finite type as an example . The general expression for the $Q$ structure in the local coordinates reads as (cf. example~\bref{ex:ODE})
\begin{equation}
Q=\theta\dl{x}+\theta V\,, \qquad V=V^\alpha(x,\psi)\dl{\psi^\alpha}
\end{equation}
Locally and under regularity assumptions one can find new coordinates $z^\alpha =z^\alpha(x,\psi)$ such that $V=\dl{z^1}$. Then in the new coordinate system $x,\theta,v^a$ where $v^1=z^1-x$ and $v^a=z^a$ for $a>1$ one finds
\begin{equation}
 Q=\theta\dl{x}
\end{equation}
so that the $Q$-bundle is locally trivial. In these coordinates equations of motion say that all $v^a(x)={\rm const}^a$. This has a clear physical meaning of time-dependent change of variables which makes the evolution trivial. If the bundle is trivial globally one concludes that the respective mechanical system is integrable.

An interesting feature which does not have a direct counterpart in the case of usual PDEs is that among gauge transformations there can be reparametrizations of the base manifold $X$. In this case under a rather general assumptions one can show that for a standard gauge PDE such that space-time reparametrizations are among its gauge symmetries one can find a local change of coordinates on the jet-bundle such that $\dh+s$ takes the form ${\rm d}_X+s_0$, with $s_0$ originating from the fiber. This was observed in~\cite{Brandt:1989et,Brandt:1993xq,Barnich:1995ap} (see also~\cite{Barnich:2010sw} for the discussion in a directly related context).  Translating this to the present language: reparameterization-invariant gauge PDE corresponds to locally-trivial $Q$-bundles.

Let us give an explicit example of a simple repara\-meterization-invariant gauge PDE and demonstrate that it is  indeed a locally trivial $Q$-bundle.  To this end consider a trivial (i.e. jet-space) ODE: independent variable is $x$ and dependent $u^i$ and there are no constraints on $u,x$. On top of this there is a ghost variable $\xi$ and $s$ is defined by
\begin{equation}
sx=s\theta=0\,, \quad su^i=\xi u^i_x\,, \quad s\xi=\xi\xi_x\,,\quad \ldots
\end{equation}
The action of $s$ on the remaining coordinates is determined by $\commut{s}{D_x}$, where $D$ denotes total derivative. It is clear that the gauge transformation determined by $s$ reads as
\begin{equation}
\delta u^i=\epsilon(x) \dl{x} u^i
\end{equation}
and indeed coincides with the transformation of $u^i$ under infinitesimal reparametrization of $x$. Here $\epsilon(x)$ denotes gauge parameter associated to ghost variable $\xi$.

The horizontal differential reads:
\begin{equation}
 \dh=\theta\left(\dl{x}+u^i_x\dl{u^i}+\xi_x\dl{\xi}+\cdots\right)
\end{equation}
Let us decompose $s$ as the sum of
\begin{equation}
 s_0=\xi\left(u^i_x\dl{u^i}+\xi_x\dl{\xi}+\cdots\right)
\end{equation}
and $s^\prime$ containing the terms proportional to $\xi_x,\xi_{xx},\ldots$. Because the structure of $s_0$ and $\dh-\theta\dl{x}$ is identical one can introduce new coordinates $\xi^\prime=\xi+\theta$ so that
\begin{equation}
Q=s+\dh=\theta\dl{x}+s_0(\xi^\prime)+s^\prime\,.
\end{equation}
Because $s_0(\xi^\prime)+s^\prime$ doesn't depend on $x,\theta$ this is precisely the product $Q$-structure.

\subsection{Parent formulation}

Gauge PDE as we defined it should be always equivalent to a standard one, i.e. to the one realized in terms of a jet-bundle.
There is a systematic way to embed a given gauge (pre)-PDE into a natural jet-bundle associated to the PDE itself.

Let us consider super jet-bundle $J^\infty(E_{T[1]X})$ associated to $\pi:E_{T[1]X} \to T[1]X$ (note that this is something like super jet-bundle of a jet-bundle).  It is again a $Q$-bundle with a total $Q$-structure being $Q^P=pr(Q)$ -- the prolongation of $Q$ from $E_{T[1]X}$ to $J^\infty(E_{T[1]X})$ and hence is a new gauge (pre)-PDE $(J^\infty(E_{T[1]X}),Q^P)$ called parent formulation.

The term `parent' is only appropriate if $E_{T[1]X}$ is itself a jet bundle because in this case among fields of the parent formulation one can find all derivatives of the original fields and hence a wide class of equivalent formulations can be obtained by equivalent reductions of the parent one.
The parent formulation was introduced in~\cite{Barnich:2010sw} (and earlier in~\cite{Barnich:2004cr} for linear system) in slightly different terms.  If the starting point equation is such that $E_{T[1]X}$ does not contain negative degree variables, i.e. it is an infinitely-prolonged equation extended by ghosts, it is more appropriate to call this formulation `intrinsic' because it is built in terms of intrinsic geometry of the equation manifold and hence doesn't depend on which jet-bundle was used to realize the equation explicitly.

We have the following:
\begin{prop}
If $(E_{T[1]X},Q)$ is a gauge pre-PDE such that it is equivalent to a standard one then its parent formulation $(J^\infty(E_{T[1]X}),Q^P)$  is equivalent to $(E_{T[1]X},Q)$.
\end{prop}
In particular, parent form of a gauge PDE is equivalent to the gauge PDE itself. The local version of the statement was formulated and proved in~\cite{Barnich:2010sw}. Proof of the global version will be given elsewhere.

Let us spell out a few corollaries:
\begin{cor}
If $(E_{T[1]X},Q)$ is locally trivial its parent formulation is of AKSZ type (i.e. is an AKSZ sigma model locally).
\end{cor}
In particular, for a reparameterization-invariant gauge PDE its parent formulation is of AKSZ type~\cite{Barnich:2010sw}.
\begin{cor}
If $(E_{T[1]X},Q)$ is a trivial $Q$-bundle defining an AKSZ model then seen as a gauge PDE $(E_{T[1]X},Q)$ is equivalent to the AKSZ model it defines.
\end{cor}

Let us make contact with the definition of the parent formulation as defined in~\cite{Barnich:2010sw}. Using the present language, suppose we are given with the standard gauge PDE $(E_{T[1]X},\dh+s)$ where $E_{T[1]X}$ is a jet-bundle over $X$ extended to the bundle over $T[1]X$. Let us consider an AKSZ sigma model with the source space $T[1]X$ and target space $(E_{T[1]X},\dh+s)$, where in the target space one takes a total degree as a grading. The resulting gauge theory was called the parameterized parent formulation because it is manifestly reparameterization-invariant and among its fields there are original independent variables $y^a,\xi^a$ (now we use different notations for coordinates on the base of $E_{T[1]X}$ to distinguish with coordinates on the source manifold). In particular $\xi^a$ give rise to ghost variables whose associated gauge transformations are precisely reparameterizations of $y^a$.

In the next step one restricts the constructed AKSZ sigma models to super maps that preserve the base space. To see that such a restriction is consistent let us restrict ourselves to local analysis and use local coordinates $y^a,\xi^a,\psi^\alpha$ in the target space. Among the fields of the AKSZ model there are all the components of $y^a(\theta)=\overset{0}{y}{}^a+\overset{1}y{}^a_b \theta^b+\overset{2}y{}^a_{bc} \theta^b \theta^c+\cdots$. Consider the infinite prolongation of the following surface in the jet-bundle of the model
\begin{equation}
\begin{aligned}
\overset{0}{y}{}^a-x^a&=0\\
\overset{k}y{}^a_{b_1\cdots b_k}&=0\,, \quad k>0\\
Q^P(\overset{0}{y}{}^a-x^a)&=0\,, \\
Q^P\overset{k}y{}^a_{b_1\cdots b_k}&=0\,, \quad k>0\,.
\end{aligned}
\end{equation}
By constriction $Q^P$ is tangent to the surface and moreover the second group of constraints implies $\overset{0}{\xi}{}^a=\theta^a, \overset{1}{\xi}{}^a_b=\delta^a_b, \overset{k}{\xi}{}^a_{b_1\cdots b_k}=0\,,k>1$ and hence the reduced system is precisely the parent formulation defined above. The BRST differential of the parent formulation can be written explicitly using the coordinates $\psi^\alpha_{a_1\ldots a_l|b_1\ldots b_k}$ and operators $d^F$, $\sigma^F$ introduced in Appendix~\bref{sec:app-sjets}. Namely, the vertical part $s^P$ of $Q^P$
reads as
\begin{equation}
 s^P={\rm d}^F-\sigma^F+\bar s\,,
\end{equation}
where $\bar s$ is the prolongation of the original BRST differential $s$.

\appendix

\section{Super jet-bundle}

\label{sec:app-sjets}
Let $E_{T[1]X}\to T[1]X$ be a $Q$-bundle. Consider the associated super (jets of super maps) jet-bundle $J^\infty (E_{T[1]X})$. Let $u^\alpha$, $x^a,\theta^a$ be local coordinates adapted to local trivialization. Basis derivatives $\dl{x^a}$ and $\dl{\theta^a}$ lift to respective total derivatives $D_a$, $D^\theta_a$ on $J^\infty (E_{T[1]X})$, which project to $\dl{x^a}$ and $\dl{\theta^a}$ by the canonical projection.  

A useful coordinate system on $J^\infty (E_{T[1]X})$ is given by $x^a,\theta^a$ along with
\begin{equation}
\begin{aligned}
u^\alpha_{a_1\ldots a_l|b_1\ldots b_k}=D_{a_1}\ldots D_{a_l}D^\theta_{b_1}\ldots D^\theta_{b_k} u^\alpha\,,\\ \gh{u^\alpha_{a_1\ldots a_l|b_1\ldots b_k}}=\gh {u^\alpha}-k\,.
\end{aligned}
\end{equation} 
Note that on $J^\infty (E_{T[1]X})$ there is also an additional grading originating from the form degree on $T[1]X$. For instance the degree of $u^\alpha_{a_1\ldots a_k|b_1\ldots b_l}$ is $-k$.

Let $Q=h^a\dl{x^a}+v^\alpha\dl{u^\alpha}$ be a vector field on $E_{T[1]X}$. Its prolongation $Q^{P}$ to super-jets is defined as follows: $Q^P=H+V$, where $H$ is horizontal, $V$ vertical and evolutionary, and $Q^P$
projects to $Q$. It follows, $H=h^aD_a$ while $V$ is determined by
\begin{equation}
\commut{D_a}{V}=0\,, \quad \commut{D^\theta_a}{V}=0\,, \qquad V \,u^\alpha=v^\alpha-h^a u^\alpha_{a|}
\end{equation} 
In particular taking $v^\alpha=0$ and $h^a=\theta^a$ one arrives at the explicit formula for prolongation $Q^P$ of ${\rm d}_X=\theta^a\dl{a}$:
\begin{equation}
\begin{aligned}
&Q^P=\theta^a\dl{x^a}+\md^F\,, 
\\
&\md^F u^\alpha_{a_1\ldots a_l|b_1\ldots b_k}=k u^\alpha_{a_1\ldots a_l [b_1|b_2\ldots b_k]}\,,
\end{aligned}
\end{equation} 
where $[]$ denote total antisymmetrization of the enclosed indices, e.g. $C_{[ab]}=\half(C_{ab}-C_{ba})$.  A useful way to work with components is to introduce generating function
\begin{equation}
 \mathbf{u}^\alpha=\sum_{k,l\geq 0}\frac{1}{k!l!}u^\alpha_{a_1\ldots a_l|b_1\ldots b_k}z^{a_1}\ldots z^{a_l}\xi^{b_k}\ldots\xi^{b_1}\,,
\end{equation} 
where we have introduced commuting auxiliary variables $z^a$ and anticommuting $\xi^a$. Vector field $d^F$ can be then defined through the following relation $\md^F\mathbf{u}^{\alpha}=\mathbf{u}^{\alpha} \dr{z^a}\xi^a$.

Let us also find an explicit expression for the prolongation $\bar\sigma$ of $\sigma=\dh-{\rm d}_X=\theta^a\Gamma(x,u)_a^\alpha \dl{u^\alpha}$ that is a nontrivial piece of the horizontal differential on $E_{T[1]X}$. For instance, for $u^\alpha_{|a}$ one gets
\begin{multline}
\bar\sigma u^\alpha_{|a}=\bar\sigma D^\theta_a u^\alpha =-D^\theta_a\bar\sigma u^\alpha=\\
-\Gamma_a^\alpha(x,u)+\theta(\ldots)=-\sigma^F u^\alpha+\theta(\ldots)\,.
\end{multline} 
This defines the vector field $\sigma^F$ which we use in the main text. It follows if the total differential has the standard structure $Q=\dh+s$  its prolongation $Q^P$ indeed coincides with the parent differential $d^F-\sigma^F+\bar s$ at $\theta=0$. 

Let us check that the parent equations of motion are precisely the $Q$-map conditions. Take a simple example
where $E_{T[1]X}$ is coordinatized by $x,\theta, w,v$ with $\gh{w}=0$, $\gh{v}=1$ and where $Q$ is given by
\begin{equation}
 Qx^a=\theta^a\,, \qquad Qw=v\,.
\end{equation} 
If $\sigma$ is a $Q$-section it follows
\begin{equation}
\label{eqqq}
 dw(x)=v_a(x)\theta^a\,, \quad d(v_a(x)\theta^a)=0\,,
\end{equation} 
where $w(x)=\sigma^*w, v^a(x)\theta^a=\sigma^*(v)$. $Q^P$ is given explicitly by
\begin{equation}
\begin{aligned}
 &Q^P x^a=\theta^a\,, \quad Q^Pw=v, \\
 &Q^P w_{|a}=w_{a|}-v_{|a}\,, \quad Q^P v_{|ab}=2v_{[a|b]}\,, \quad \ldots
\end{aligned}
\end{equation} 
The body of the zero locus of $Q^P$ is the stationary surface and indeed we get conditions $w_{a|}-v_{|a}=0$ and 
$v_{[a|b]}=0$ in accord with~\eqref{eqqq}.

The explicit form of the complete $Q^P$ simplifies if one uses coordinates $\psi^\alpha_{a_1\ldots a_l|b_1\ldots b_k}$ that coincide with $u^\alpha_{a_1\ldots a_l|b_1\ldots b_k}$ at $\theta=0$ and satisfy $D^\theta_a \psi^\alpha_{a_1\ldots a_l|b_1\ldots b_k}=0$. In these coordinates one finds that $V \psi^\alpha_{a_1\ldots a_l|b_1\ldots b_k}$ is $\theta$-independent and the expression for $(\dh+s)^P$ takes the form $D_h+\md^F-\sigma^F+\bar s$, where $D_h=\theta^a D_a$


\end{document}